\documentclass[%
reprint,
superscriptaddress,
amsmath,amssymb,
aps,
pra,
]{revtex4-2}
\usepackage{graphicx}
\usepackage{dcolumn}
\usepackage{bm}
\usepackage{amsmath}
\usepackage{hyperref}
\usepackage{tikz-feynman}

\begin{document}
	
	\preprint{APS/123-QED}
	
	\title{Perturbative approaches to quantum field theory in curved space-time}
	\author{Jesse Huhtala}%
	\email{jejohuh@utu.fi}
	\affiliation{Department of Physics and Astronomy, University of Turku, 20014 Turku, Finland}%
	
	\author{Iiro Vilja}
	\email{vilja@utu.fi}
 	\affiliation{Department of Physics and Astronomy, University of Turku, 20014 Turku, Finland}

	\author{Nicola Lo Gullo}
	\email{nicolino.logullo@vtt.fi}
	\affiliation{Quantum algorithms and software,VTT Technical Research Centre of Finland Ltd}
    \affiliation{Dipartimento di Fisica, Universit\`a della Calabria, 87036 Arcavacata di Rende (CS), Italy}

	\date{\today}
	
	\begin{abstract}
		Despite the large amount of work done in quantum field theory in curved space-times, there are not great many results available for perturbative calculations of particle processes in these systems. Such processes are expected to be important in the early stages of the universe, as well as near highly relativistic objects like black holes and, recently, in effective field theories of condensed matter systems. The difficulties with carrying out perturbative calculations in curved space-times are related to the practical difficulty of quantizing in curved space-times. This suggests the need for investigating new approximations and comparing the results of different methods of carrying out the calculations. In this paper, we compare different perturbative approaches to particle scattering problems and illustrate them with examples. In particular, we look at a 1+1 dimensional static spacetime. We also examine the spacetime particle generation in a 1+1 Robertson-Walker universe, where we find that the curved space LSZ formula yields new information compared to the in-in formalism.

	\end{abstract}
	
	\maketitle
	
	
	\section{\label{sec:intro}Introduction}
	
	By now, quantum field theory in curved space-times is a well-established subject, with several textbooks devoted to it (e.g. the classic \cite{Birrell1982}, or the more recent \cite{Mukhanov2007}). There is a healthy community of researchers working on the topic \footnote{A fairly comprehensive list of people in the field can be found from the speakers and participants of the recent Quantum Field Theory in Curved Spacetimes Workshop at \url{https://sites.google.com/view/qftcsworkshop/}(referenced on 10th November, 2022)}, with plenty of recent progress on, for example, the mathematical foundations of quantum field theory (QFT) in curved space-times. 
	
	On the mathematical side, work has largely concentrated on the algebraic QFT framework, which has previously been used to formalize QFT in flat space-time in a mathematically rigorous way. There are various existence proofs for the physically relevant quantities in QFT in curved space-times, and rigorous algebraic formalization seems to be proceeding apace \cite{Fewster2021, Fewster2020, Hollands2001, Hollands2002, Hollands2010, Rejzner2021}. 
	
	Recently, it has also become evident that certain open quantum systems can be treated as being in effectively curved space-times. A small open quantum system that can be described by a non-Hermitian Hamiltonian might be modeled by placing the system on a curved surface, thus potentially expanding the use of the curved space-time formalism to ordinary quantum systems \cite{Lv2022}. Quantum technologies have also become sufficiently sensitive that ground-based experiments may soon be possible \cite{Howl2019}. There is also recent work on effectively curved spacetimes in Bose-Einstein condensates \cite{Viermann2022}. Hence, even experimental realization of some curved space-time systems seems now to be within reach.
	
	Despite these successes, there are various practical problems with carrying out curved space-time calculations. As was already pointed out by Fulling in 1973 \cite{Fulling1973}, quantization in curved space-times faces several conceptual and technical problems --- even the notion of "particle"\ seems elusive.  In general globally hyperbolic space-times, it is difficult to even separate the solution in to positive and negative energy parts, which is necessary for canonical quantization \cite{Magnon1975}. Despite of the aforementioned mathematical progress in algebraic QFT, the methods developed do not easily lend themselves to practical calculation. The fundamental reason is the inequivalence of the in and out vacuums that complicates matters. Therefore the calculation of S-matrix elements (or other quantities such as in-in expectation values) is an arduous task.
 
 Though there are enormous practical difficulties, some methods and results do exist. For example, the LSZ reduction formula can be generalized to curved space-times \cite{Birrell1979}. The added-up formalism, which attempts to circumvent the problems associated with inequivalent vacuums \cite{Audretsch1985,Audretsch1987}  has been applied to various cosmological systems \cite{Lankinen2017,Lankinen2018a,Lankinen2018b}. Another method of circumventing the vacuum issue is the Schwinger-Keldysh formalism \cite{Jordan1986}. However, results have only been obtained in a restricted class of space-times, and with somewhat dubious methods (for example, in the case of the FRW metric, one is forced to use "instantaneous quantization", whereby each instant of time essentially requires a new quantization procedure \cite{Mukhanov2007}); in the case of the Schwinger-Keldysh formalism, the calculations tend to become very complicated very quickly, as can be seen already in the example in \cite{Jordan1986}. 
  
Various results are available for the backreaction of quantum fields on the curved space-time, especially for the stress-energy tensor; a few examples from recent years can be found in \cite{Taylor2021, Zilberman2021, Meda2021a, Pla2021, Meda2021b, Bernardo2021}. There is, however, an evident need for more effective methods in carrying out practical perturbative calculations in curved space-times; it is still desirable to obtain more results.
	
The previously mentioned ground-based experiments and effectively curved space-times are potential application areas, but so are the cosmological phenomena. Near the time of the Big Bang the effect of the curvature of space-time on particle processes was likely significant \cite{Birrell1982, Lankinen2017}, and cosmological measurements are performed with ever-increasing accuracy, necessitating the inclusion of more processes in to the cosmological particle models. However, these applications require specific calculational methods different from the backreaction calculations.
	
So, to that end, we compare the results of a perturbative LSZ-type calculation \cite{Birrell1979} and "added-up probability"\ calculation \cite{Audretsch1985,Audretsch1987} with each other, and discuss their equivalence in various scenarios. To our knowledge, there aren't any other examples of the curved space LSZ formula in \cite{Birrell1979} being used for a concrete perturbative calculation in the way that we do here (though see \cite{Haba2006}), so we hope to illuminate its use and potential benefits.  We provide examples of doing such calculations, and arguments for and against utilizing in-out amplitudes.
	
	The paper is organized as follows. In section \ref{sec:addedup}, we review the added up probability formalism of Audretsch and Spangehl. A similar review follows in \ref{sec:lsz} for the curved space Lehmann-Symanzik-Zimmermann (LSZ) reduction formula. Section \ref{sec:differences} analyzes the differences in the LSZ and added up formalisms. Section \ref{sec:results} presents results for two models. In section \ref{sec:discussion}, we discuss the results and their meaning for perturbation theory in curved space-times.
	
	\section{\label{sec:addedup}In-in probability amplitudes}
	In-in calculations are based on the idea that in-out matrix elements, such as
	\begin{align}
		S_{\text{in-out}} = \langle k_i,\text{out}| p_i, \text{in}\rangle
	\end{align}
	can be hard to both interpret and calculate in curved space-times due to the fact that, for example, the in-out effective field equations require the specification of end-point boundary conditions \cite{Jordan1986}. On the other hand the in-in amplitudes
	\begin{align}
	S_{\text{in-in}} = \langle k_i, \text{in}| p_i,\text{in}\rangle
	\end{align}
	can be interpreted as expectation values and utilized perturbative techniques. Hence it is often argued that it is more physically illuminating to calculate in-in matrix elements \cite{Jordan1986, Mukhanov2007, Audretsch1985, Audretsch1987}.
	
	One example of in-in methods is the added-up probability method that was developed by Audretsch and Spangehl \cite{Audretsch1985, Audretsch1987}.  The basic idea is that, in curved space-time calculations, the space-time itself can be a source of particle generation. However, in the case of conformal coupling, the background space-time does not appear in the equations of motion for massless particles. Hence, they are good indicators of the outcomes of particle processes, as they won't interact directly with the space-time.
	
	Mathematically the idea can be realized in two scalar field system by writing the S-matrix element in curved spacetime as
	\begin{align}
		&\langle \mathrm{out}, d^\phi s^\psi |S |c^\phi r^\psi ,\mathrm{in}\rangle \nonumber\\
		&= \sum _{g,t} \langle \mathrm{out},d^\phi s^\psi |g^\phi t^\psi , \mathrm{in}\rangle \langle \mathrm{in},g^\phi t^\psi| S | d^\phi s^\psi, \mathrm{in}\rangle  ,
	\end{align}
	where the sum $g,t$ ranges over a complete basis of particle states, such that $\sum _{g,t}\langle g^\phi t^\psi , \mathrm{in}\rangle \langle \mathrm{in},g^\phi t^\psi| = I$. 
 We indicate massive, non-conformally coupled particles by $\phi$ and massless, conformally coupled particles by $\psi$. The first factor on the right hand side is proportional to $\langle \mathrm{out}, 0|0,\mathrm{in}\rangle  $, as can be seen by a direct calculation (\cite{Audretsch1985}, eqs 3.1-3.3), with the proportionality factor calculated from the Bogoljubov coefficients. In flat space-time, this term would be unity, but in curved space-times, the in- and out vacuums may differ from one another. However, we are mostly interested in the particles that take part in processes, instead of the ones spontaneously generated by the space-time: as stated above, massless particles with conformal coupling are not generated by the space-time. This makes them good counters of particle processes.
	
	Thus, we wish to ask the question: what is the probability for the massless particles to end up in a particular state, regardless of what the massive particles do? In other words, we want to calculate the added-up transition probability \cite{Audretsch1985, Audretsch1987}
	\begin{align}
		 w^{add}(s^\psi |c^\phi r^\psi) = \sum _d |\langle \mathrm{in},d^\phi s^\psi | S | c^\phi r^\psi ,\mathrm{in}\rangle |^2, \label{eq:addedupform}
	\end{align}
	where the sum runs over all massive particle states. Thus one can perform perturbation theory on the S-matrix as per usual setting 
	\begin{align}
		S = 1 - i\lambda A + \mathcal{O}(\lambda ^2),
	\end{align}
	calculating the reduced matrix element, and so on; from this point on, the calculation proceeds exactly as in flat space-time, only the mode functions are changed. Note, however, that the interaction term in the curved space contains the metric determinant, i.e. for a flat space-time interaction $\mathcal{L}_{\text{int}}$, in curved space-time you end up with the regulated S-matrix
	\begin{align}
	S = \lim _{\alpha \rightarrow 0}\exp \bigg[ i\int d^Dx\sqrt{-g} \mathcal{L}_{\text{int}} \exp (-\alpha |t|) \bigg] .
	\end{align}

	Essentially, eq.\ \eqref{eq:addedupform} discards the significance of the Bogoljubov coefficients on the end result as the formula is relying on the conformal coupling of the massless field, and we have to make an informed choice about what information we consider physically relevant.
	
	Another example of in-in methods is the Schwinger-Keldysh formalism \cite{Schwinger1961, Jordan1986}. The idea is to set the time integration in the action on a contour consisting of one part going forwards in time and one part going backwards in time. These different parts of the contour are then given different dynamics by setting the sources as $J_+$ and $J_-$ for the forwards and the backwards parts, respectively. One proceeds by inserting the set of complete out-states \cite{Jordan1986}:
	\begin{align}
		e^{iW[J_+,J_-]} = \sum _\alpha \langle \text{in},0|\text{out},\alpha \rangle _{J_-}\langle \text{out}, \alpha | \text{in},0\rangle _{J_+} .
	\end{align}
	 Then, in the path integral formalism, we get
	\begin{align}
		&e^{iW[J_+,J_-]} \nonumber\\
		&\propto \cdot \int D\phi ' \int e^{-i(S[\phi _-] + \int d^4xJ_-\phi _-)} \mu [\phi _-] d\phi _-\nonumber\\
		&\times \int e^{i(S[\phi _+] +  \int d^4xJ_+\phi _+)} \mu [\phi _+]d\phi _+.
	\end{align}
	where the functional integral over $\phi '$ takes care of a potentially complicated boundary condition $\phi _+ = \phi _- = \phi$ on $\Sigma$, with $\Sigma$ the space-like hypersurface that defines an instant of time which serves as the turnaround point for the integration contour. The integral measure terms $\mu$ have to be chosen carefully, and aren't guaranteed to be the same as in flat space; there is some discussion of this in various places, for instance, in \cite{Jordan1986}. 
	
	With this setup the Green's functions are generated by taking functional derivatives with respect to $J_+$ and $J_-$. Finally the sources are set to equal and then zero, as in the usual path integral formalism, but different combinations still yield different propagators: two derivatives with respect to $J_+$ yield the time ordered 2-point propagator, two derivatives with respect to $J_-$ give the anti-time ordered propagator, and the mixed derivatives give no time ordering; more details can be found in \cite{Jordan1986}.
 
	\section{\label{sec:lsz}Curved spacetime LSZ reduction formula}
 
	It is possible to derive reduction formulae analogous to those of flat space-times also for curved backgrounds. This was originally done by Birrell and Taylor in \cite{Birrell1979}. We note that the primary difference in the derivation of the flat and curved space-time reduction formulae is that one cannot use simple relation $|\text{in},0\rangle = |\text{out},0\rangle$. This means non-trivial Bogoljubov coefficients are present --- since the creation and annihilation operators for the out and in vacuums are different, one obtains extra terms in the reduction formula. 
	
	We simply quote the end result of a lengthy derivation; for one species of scalar field the matrix element reads as \cite{Birrell1979}
	\begin{align}
		&\langle \text{out}, \{k_i\}^{(s)}|\text{in},\{p_i\}^{(r)}\rangle = \sum _\rho i^{(r-l)/2} \nonumber \\
		&\times \bigg[O(z_{\rho ( 1)})\cdots O(z_{\rho (k)})\alpha ^{-1}_{m_{\rho(k+1)}n_{\rho (k+1)}}\cdots \alpha ^{-1}_{m_{\rho (l)}n_{\rho (l)}} \nonumber \\
		&\times \Lambda _{\rho (l+1) \rho (l+2)}\cdots \Lambda _{\rho (r-1) \rho (r)} \sum _\sigma i^{(s-l-\omega +k )/ 2} \nonumber \\
		&\times Q_{z_{\sigma (1)}}\cdots Q_{z_{\sigma (\omega )}} V_{\sigma (\omega +1) \sigma (\omega +2)} \cdots V_{\sigma (s-l+k-1) \sigma (s-l+k)} \nonumber \\
		&\langle \text{out},0|T(\phi (z_{\sigma (1)}) \cdots \phi (z_{\sigma (\omega )}) \phi (z_{\rho (1)}) \cdots \phi (z_{\rho (k)})) |\text{in},0\rangle\bigg] . \label{eq:lszred}
	\end{align}
Here, the sums are understood to be over permutations of indices; $O_z = i\sum _p \alpha _{m,p}^{-1}\int \sqrt{-g_z}d^4z f^1_{p}(z)$ is an integral operator for incoming particle with mode functions $f^1$, and $Q_z$ is the corresponding operator for outgoing particle with mode function $f^2$. The numbers $s$ and $r$ indicate how many particles are present in the sets $\{k_i\}^{(s)}$ and $\{p_i\}^{(r)}$, $\alpha$ and $\beta$ are Bogoljubov coefficients, and $V$ and $\Lambda$ are combinations of Bogoljubov coefficients given by $\Lambda _{ij} = -i \sum _p \beta _{pm_j}\alpha _{m_ip}^{-1} $ and $V_{ij} = i\sum _p \beta _{n_j p}\alpha _{pn_i}^{-1}$. We define $\alpha$ and $\beta$ in the usual fashion: $\alpha _{ij} = (\overline{u}_i,u_j)$ and $\beta _{ij}=-(\overline{u}_i,u_j^*)$, with $u_i$ the mode functions in the in-vacuum, $\overline{u}$ the mode functions in the out-vacuum both of them forming complete sets.
	
	The formula is rather opaque. It is easier to understand it by showing how to reduce just one particle from the in and out-states. For  an in-states this is given by
	\begin{align}
		&\langle \text{out}, \{k\}^{(s)} | \text{in}, \{p\}^{(r)}\rangle \nonumber \\
		&=O_{z_i}\langle \text{out},\{k\}^{(s)}|T\phi (z_i)|\text{in},\{p\}^{(r)}-\{p_i\}\rangle \nonumber \\
		&+\sum _{j=1}^s \alpha ^{-1}_{m_in_j}\langle \text{out},\{k\}^{(s)}-\{k_j\}|\text{in}, ,\{p\}^{(r)}-\{p_i\}\rangle \nonumber \\
		&+i\sum _{j=1}^{r-1} \Lambda _{ij}\langle \text{out},k^{(s)}|\text{out},\{p\}^{(r)}-\{p_i,\, p_j\}\rangle\,, \label{eq:redright} 
	\end{align} 
and for the out-state, the reduction works as
	\begin{align}
		&\langle \text{out},\{k\}^{(s)} | \text{in}, \{p\}^{(r)}\rangle \nonumber\\
		&=Q_{z_i}\langle \text{out}, \{k\}^{(s)}-\{k_i\}|T\phi (z_i) |\text{in},\{p\}^{(r)}\rangle \nonumber \\
		&+ i\sum _{j=1}^s V_{ij}\langle \text{out}, \{k\}^{(s)}-\{k_i,\,k_j\}|\text{in}, \{p\}^{(r)}\rangle . \label{eq:redleft}
	\end{align}
More details are found in \cite{Birrell1979}. 

In practice, it is easier to use forms \eqref{eq:redleft} and \eqref{eq:redright} rather than \eqref{eq:lszred}, since it is hard to keep track of the various combinations and permutations involved in \eqref{eq:lszred}. It is worth noting that the reduction formula differs from the flat space-time formula by the introduction of new terms dependent on the Bogoljubov coefficients $V_{ij}$, $\Lambda _{ij}$ and $\alpha ^{-1}_{ij}$. Setting $\alpha = I$ and $\beta = 0$, and assuming that there are no "straight-through" particles that have the same momentum in the in and out states, we get the flat space formula. 
	
 Also, in static space-times $\alpha = I$ and $\beta = 0$, but, since momentum is not necessarily conserved, one might expect there to be particles with the same four momentum in the in and out states even in the presence of interactions. However, when calculating physical quantities such as decay rates etc., such scenarios represent a set with a zero measure. So, as long as we are not interested in the in-out matrix elements per say but rather observables calculable from them, we can discard the term in \eqref{eq:redright} involving $\alpha ^{-1}$.
	
	\section{\label{sec:differences} The differences between the in-out and the in-in formalism}
	\subsection{\label{sec:static}Equivalence of the in-out and in-in formalisms in static spacetime}
	Let us start by proving that the in-in and in-out LSZ formalisms presented in the previous sections are in fact equivalent in static spacetimes. This seems intuitive, since in static spacetimes the in and out vacuums are the same; it is also manifestly true if we calculate in-out amplitudes by setting $\langle k_i, \text{out} | p_i, \text{in}\rangle = \sum _m \langle k_i,\text{out} | k_m, \text{in}\rangle \langle k_m, \text{in}|p_i, \text{in}\rangle $, since then the Bogoljubov factor $\langle k_i,\text{out} | k_m, \text{in}\rangle$ is just 1. We should therefore be greatly distressed if it turned out the LSZ formula wasn't equal to the in-in formalism in static spacetimes. We will prove this for two fields --- the generalization is easy to see.
	
	Let $\phi ^1, \phi ^2$ be the fields associated with the system having a static spacetime in D dimensions, with $\phi ^2$ the outgoing and $\phi ^1$ the incoming fields, and $a^1,a^2$ the associated annihilation operators. Let the mode functions be called $f^1$ and $f^2$ respectively. The interaction is then assumed to be of the form $\lambda (\phi ^1)^z(\phi ^2)^x $. In that case, the added up formalism gives the series
	\begin{align}
		&\langle k_1...k_m|S|p_1..p_l\rangle \nonumber \\
		&= \sum _{n=0}^\infty \frac{i^n \lambda }{n!} \langle k_1...k_m|\prod _{k=0}^n \int  d^Dz_k \sqrt{-g_{z_k}}\mathcal{L}_I(z_k)|p_1..p_l\rangle \\
		&= \sum _{n=0}^\infty \frac{i^n \lambda ^n }{n!} \langle 0|(a^2_{k_1})...(a^2_{k_m})\prod _{k=0}^n \int  dz_k \nonumber \\
		&\times \sqrt{-g_{z_k}}\mathcal{L}_I(z_k)(a^1_{p_1})^\dagger...(a^1_{p_l})^\dagger |0\rangle \\
		&= \sum _{n=0}^\infty \frac{i^n \lambda ^n}{n!} \langle 0|\int d\tilde{k}_1'...d\tilde{p}_{o}' [a^2_{k_1},(a^2_{k_1'})^\dagger]...[a^2_{k_i},(a^2_{k_i'})^\dagger ]\nonumber \\
		&\times [(a^1_{p_1'}),(a^1_{p_1})^\dagger]...[(a^1_{p_o'}) , (a^1_{p_o})^\dagger] \nonumber \\
		&\times \prod _{k\in \mathbb{L}} \int  d^Dz_k \sqrt{-g_{z_k}}(f^2_{k_1'}(z_k))^*..f^1_{p_o'}(z_k) |0\rangle . \label{eq:seriesformadded}
	\end{align}
Here, we understand $\mathbb{L}$ to be the set of such $n$ that the vacuum is not annihilated. Also, we take the measures $\tilde{k}_i$ to contain both the discrete and continuous part of the spectra.
	
	The annihilation and creation operators have been replaced with commutators, since $aa^\dagger |0\rangle  = [a,a^\dagger]|0\rangle $. Note that in \eqref{eq:seriesformadded}, there are also commutators that are combinations of internal vertices only. The commutators simply produce Dirac deltas and some terms related to normalization, which ought to be canceled by the integrals over the spectrum; hence we can simply do those integrals and end up with
	\begin{align}
		&\langle k_1...k_m|S|p_1..p_l\rangle = \sum _{n=0}^\infty \frac{i^n \lambda ^n}{n!}  \nonumber \\
		&\sum _{\sigma}N_\sigma  \prod _{k\in \mathbb{L}} \int  d^Dz_k \sqrt{-g_{z_k}}(f^2_{k_1}(z_k))^*\cdots f^1_{p_o }(z_k) . \label{eq:finaladded}
	\end{align} 
	Here we take the sum over $\sigma$ to be a sum over all the different Feynman diagrams arising from \eqref{eq:seriesformadded}, each of course having its own combinatorial factor $N_\sigma$. Again, note that this expression can contain loop terms, and in the general case we do not assume that the series is renormalizable.
	
	We now compare to the in-out formalism. First, we note that the reduction formula is much simplified in the static case; it reads
	\begin{align}
		&\langle k_1 ... k_m |  p_1...p_l\rangle = \prod _{n=0,j=0}^{m,l} O_{x_j}Q_{y_n}\nonumber \\
		&\times \langle \text{out},0|T\phi ^2 (y_1)...\phi ^2(y_m)\phi ^1(x_1)...\phi ^1(x_l)|\text{in},0\rangle   .
	\end{align}
	
	  Starting off from this formula, we get immediately
	\begin{align}
		&\langle k_1...k_m|p_1...p_l \rangle \\
		&= \prod _{n=1,j=1}^{m,l} \int d^Dy_nd^Dx_j \sqrt{(-g_{y_n})(-g_{x_j})}\nonumber\\
		&\times (f^2(y_n))^*f^1(x_j)\langle 0 | T(\phi ^2 (y_n)\phi ^1(x_j))|0\rangle \label{smgen} \\
		&= \prod _{n=1,j=1}^{m,l} \int dy_ndx_j \sqrt{(-g_{y_n})(-g_{x_j})}(f^2(y_n))^*f^1(x_j)\mathcal{K}_{y_n}\mathcal{K}_{x_j}\nonumber \\
		&\times \sum _{b=0}^\infty \frac{i^b\lambda ^b}{b!} \int \prod _{v=0}^b d^Dz_{v}\sqrt{-g_{z_v}}\langle 0 | T(\phi _0 ^2 (y_n)\phi _0 ^1(x_j))\mathcal{L}_{I}(z_v)|0\rangle .
	\end{align} 
	Note that we have written an equals sign to the series expansion, even though it has been argued that many series in QFT are asymptotic \cite{Dyson1952}. Hence the equality is only formal. We can now apply the operators $\mathcal{K}$ recursively in to the expectation value of the free fields. Let us denote them with $\tau$, as previously. Then we can see that
	\begin{align}
		&\prod _{n,j,v}\mathcal{K}_{y_n}\mathcal{K}_{x_j}\tau (y_n,x_j,z_v) \nonumber \\
		&= \prod _{n,j}\bigg( \frac{-i}{\sqrt{-g_{x_j}}}\bigg)\bigg( \frac{-i}{\sqrt{-g_{y_n}}}\bigg) \sum _\sigma \delta (x_{\sigma _{1}}-z_{\sigma _{1}})\delta (y_{\sigma _{4}}-z_{\sigma _{4}}) .\label{eq:intermediate} 
	\end{align}
	The permutations $\sigma$ are over all the possible ways to connect the different indices, not including combinations of external vertices such as $\delta (x_1-x_2)$, $\delta (y_1-y_2)$ and so on. It may not seem immediately obvious that such terms aren't allowed from this form; showing why this is the case is illustrative. Whenever one ends up with a Dirac delta connecting two external vertices, one necessarily ends up being unable to do one of the integrals over the external vertices:
	\begin{align}
		\int d^Dy_i\sqrt{-g_{x_i}}d^Dx_j(f^2(y_i))^*f^1(x_i)\mathcal{K}_{x_i}\delta (y_i-x_i).
	\end{align}
Due to the properties of the Dirac delta function, this is equivalent to
	\begin{align}
		-\int d\vec{y_i}\sqrt{-g_{x_i}}d\vec{x_j}(f^2(y_i))^*(\mathcal{K}_{x_i}f^1(x_i))\delta (y_i-x_i).
	\end{align}
	But according to definition, $\mathcal{K}_{x_i}f^1(x_i) = 0$. So all such terms end up vanishing; this is true regardless of whether the metric is static or not --- this observation does not even require the equation of motion to be separable to space and time dependent parts (although in such cases canonical quantization tends to be impossible).
	
	Now we note that the metric terms are cancelled immediately by the prefactors in \eqref{eq:intermediate}, and since they produce Dirac delta functions, we may immediately perform the $x$ and $y$ integrations, leaving us with
	\begin{align}
		&\langle k_1...k_m|p_1...p_l \rangle = \prod _{n,j}^{m,l} \sum _{b=0}^\infty \nonumber \\
		&\times \sum _\sigma N_\sigma  \frac{(-i)^{l+m}i^b\lambda ^b}{b!}  \int \prod _{v=0}^{b-n_l} dz_{v}\sqrt{-g_{z_v}} (f^2(z_v))^*f^1(z_v) \label{eq:finallsz}
	\end{align}
	where $n_l$ is the number of loops. This is of the same form as with the in-in formalism. The only question that remains is if the set $\mathbb{L}$ of non-vacuum-annihilating terms produces the same terms as the LSZ perturbation series. The question essentially reduces to asking whether, after applying the differential operators $\mathcal{K}$, the remaining free field Green functions $\tau (z_1,...,z_o)$ correspond to the same terms as you get from the in-in formalism. Indeed they do: in the case of static spacetimes, this recursive formula is equivalent to using Wick's theorem \cite{Birrell1979}, since there is no trouble with Wick's theorem when the in and out vacuums are the same, and in turn the "in-in" formalism in static spacetimes works exactly as in the flat case. Since there is no change from flat space in either formalism, except in the mode functions, they yield the same terms with the same combinatorial factors just as they do in flat spacetime. In this respect, there is no difference to flat spacetime whatsoever. Hence these two expressions \eqref{eq:finaladded} and \eqref{eq:finallsz} are equivalent, except possibly for a global phase, which is physically insignificant.
	
	Also, in the added up formalism one sums over all the out states with massive particles. However, this generalization is trivial, since it is a sum over terms like \eqref{eq:finallsz} and works exactly the same way as the calculation for two fields done here. Whether this is a sensible thing to do in the in-out case will be discussed later, when we give examples.
	
	It is worth noting that the LSZ formula and the in-in formalism are of course equivalent in the flat spacetime case. However, even for the case of a static spacetime, the result is not that easy to see directly from the LSZ formula. The presence of a non-trivial metric forces the equations in to a more complicated form. These ideas should become clearer with the examples that follow.
	
	\subsection{\label{sec:discinout} In-out amplitudes in non-static spacetimes}
	
	The equivalence of the previous section does not hold for more complicated spacetimes; it obviously does not hold for time-dependent metrics, where spontaneous (massive) particle creation is possible. Nevertheless, the in-in formalism is frequently applied to such problems, and not without reason. A similar calculation as above in fact also applies for particle processes in which the out-state contains only conformally coupled massless particles.
	
	For concreteness, consider $\langle k_1k_2|p_1\rangle$, where $k_i$ are associated with a conformally coupled massless field and $p_1$ is a massive particle. Associate the operators $Q, \psi $ to the massless particles, $O, \phi $ to the massive particle. The LSZ formula yields the result
	\begin{align}
		\langle k_1 k_2|p_1\rangle =& Q_{y_2}Q_{y_1}O_{x_1} \langle 0, \text{out}|T\psi (y_2) \psi (y_1) \phi (x_1)|0,\text{in}\rangle \nonumber \\
		&+ O_{x_1}iV_{k_1k_2}\langle 0,\text{out}|\phi (x_1)|0\rangle .
	\end{align} 

	The second term looks curious, and certainly differs from the in-in amplitude. But since it depends on Bogoljubov factors $\beta$, and since those are zero for a conformally coupled massless field, the term is zero. The argument easily generalizes to any number of conformally coupled particles in the out-state and massive particles in the in-state. Incidentally, this means that the cosmological in-in calculations in \cite{Lankinen2017, Lankinen2018a, Lankinen2018b} involving a decay process in to massless particles are genuinely unaffected by the time-dependence of the metric --- it only comes in to play when finding the mode functions, but the in-out calculation would produce the exact same result as the in-in calculation. This is the same result as the assertion in \cite{Audretsch1985}; the LSZ formula offers some concreteness to the argument presented there.
	
	We will return to the time-dependent case in Section \ref{sec:quartet}, when we look at some terms in the energy density of a Robertson-Walker metric.
	
	\section{\label{sec:results}Case studies}
	We will now work out two particular cases to make the previous discussion concrete. The first is a static space-time calculation to illustrate the use of the LSZ formula for extracting a perturbation series. We compare it also to the in-in calculation.
	
	We then do a calculation for the Robertson-Walker metric, emphasizing certain aspects of particle creation that are opaque in the in-in formalism. In particular, it is possible to extract information about the number of particle pairs generated by the spacetime.
	
\subsection{\label{sec:2ax} 1+1 dimensional case, $g = 2ax\eta $}
	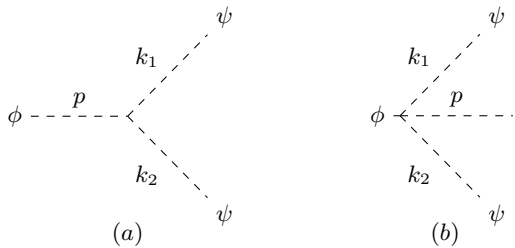
\begin{figure}
	\begin{tikzpicture}
	\begin{feynman}
		\vertex (a) {$\phi $};
		\vertex [right=of a] (b);
		\vertex [below=1.3cm of b] {$(a)$};
		\vertex [above right=of b] (f1) {$\psi $};
		\vertex [below right=of b] (f2) {$\psi $};
		\diagram* {
			(a) -- [scalar, edge label=$p$] (b) -- [scalar, edge label=$k_1$] (f1),
			(b) -- [scalar, edge label'=$k_2$] (f2),
		};

	\end{feynman}

	\end{tikzpicture}\hspace{1.5cm}
	\begin{tikzpicture}
	\begin{feynman}
		\vertex (a) {$\phi $};
		\vertex [right=0.3cm of a] (b);
		\vertex [right of=b] (f3);
		\vertex [] at (0.9,-1.58) {$(b)$};
		\vertex [above right=of b] (f1) {$\psi $};
		\vertex [below right=of b] (f2) {$\psi $};
		\diagram* {
			(a) -- [scalar]  (b) -- [scalar, edge label=$p$] (f3),
			(b) -- [scalar, edge label'=$k_2$] (f2),
			(b) -- [scalar, edge label=$k_1$] (f1),
		};
		
	\end{feynman}
	
	\end{tikzpicture}\hfill 
\caption{The possible first order Feynman diagrams for the decay of a massive particle in to two massless ones. Note, that generally neither diagram (a) nor (b) conserves momentum; obvioulsly the latter diagram  never conserves energy, since a particle is spontaneously created by the background. In our metric, which is static, there is no background-induced particle creation, so (b) does not contribute.}
\label{fig:diagram}
	\end{figure}

	The model we will be doing the calculations for has the metric $g_{\mu \nu} = 2ax\eta _{\mu \nu}$, which we solve for $x\geq 0$. The metric is conformally flat. We will calculate the decay channel $\phi \rightarrow \psi \psi$, where $\phi$ is a massive scalar and $\psi$ is a massless one. Our aim is to calculate the decay rate $\Gamma _{\phi \rightarrow \psi \psi}$. The diagrams up to first order can be seen in Figure \ref{fig:diagram}.
	
	The Lagrangian is
	\begin{align}
		L &= \int d^2x \sqrt{-g}\bigg[ \mathcal{L}_{\phi ,\text{free}} + \mathcal{L}_{\psi , \text{free}} + \mathcal{L}_{\text{int}} \bigg],\\
		\mathcal{L}_{\text{int}} &= -\lambda \phi \psi ^2.
	\end{align}
	The free particle Lagrangians are the same as in flat space-time, and the metric is the one given above.
	
	For non-interacting fields we have two different equations of motion. In the massless case, we end up with the same exponential modes as for a flat space-time. For the massive field, we are lead to the mode equation
	\begin{align}
		\bigg( -\frac{d^2}{dx^2} - \omega ^2 + 2axm^2 \bigg) \chi (x) = 0 \label{modeequation}
	\end{align}
	from which, given the boundary condition $\chi (0) = 0$, we get the normalized (in accord with \cite{Fulling1973}) eigenfunctions
	\begin{align}
		\chi_n (x) &=	\mathcal{N}\text{Ai}\bigg( \frac{-\omega_n ^2+2 a m^2 x}{2^{2/3} \left(a m^2\right)^{2/3}} \bigg) \label{modesol},\\
		\mathcal{N} &= \sqrt{\frac{2am^2}{2^{2/3}(am^2)^{2/3} \text{Ai}'(\zeta _n)^2}} .
	\end{align}
	Here, Ai is the Airy function and $\zeta _n$ are its zeros \cite{Stegun1964}, which all are negative. The inner function of Ai in Eq. \eqref{modesol} we denote by $f(x, \omega)$. In our case, because of the boundary condition we have chosen, the energy eigenvalues are given by
 \begin{equation}\label{dispersion}
     \omega ^2 _n= - {2^{2/3}(am^2)^{2/3}}\zeta _n \,.
 \end{equation} 
 It is worth noting, that our calculation differs from \cite{Fulling1973} because of this boundary condition: we in effect insist that particles do not "leak out" of the universe defined by $x\geq 0$. This results in a discrete spectrum labeled by positive integer $n$.

	The solution \eqref{modesol} leads to a field operator of the form
	\begin{align}
		\phi (x) = \sum _{n}  \mathcal{N} \text{Ai}\bigg( \frac{-\omega_n ^2+2 a m^2 x}{2^{2/3} \left(a m^2\right)^{2/3}} \bigg) \bigg[ e^{-i\omega_n t}a^\dagger + e^{i\omega_n t}a \bigg] \label{phi},
	\end{align}
	which are to be used in our calculations. We will also denote the differential operator, the Klein-Gordon operator, in \eqref{modeequation} by $K_x $.
 
	\subsubsection{\label{ssec:addedupresult} In-in probability amplitude}
	
	Using field operator \eqref{phi}, we want to calculate $|\langle \psi _{k_1}\psi _{k_2}|S|\phi _p \rangle |^2$; this is the tricky part of obtaining the decay rate. As the appearing time integral results in a Dirac delta, to first order we end up with the matrix element
	\begin{align}
		S_{k_1k_2}^p = &-i2\lambda \delta (\omega _{k_1}+\omega _{k_2}-\omega _p)\nonumber\\
		&\times  \int dx\bigg[ \chi (x) \psi_s ^2(x)\sqrt{-g} \bigg]\label{reduced}\,,
	\end{align}
where $\omega_p$ is the incoming particle energy having possible values given by Eq. \eqref{dispersion}, and $\psi_s (x)$ is the spatial part of the flat-space Klein-Gordon solution.

	After rather standard technical calculations (some details in Appendix \ref{app:addedup}), we find
	\begin{align}
		\Gamma _{\phi \rightarrow \psi \psi}=\frac{16\pi ^3 a \lambda ^2}{\omega _p^2 m^2 \mathrm{Ai}'(\zeta _n)^2}I_n \label{finaladdedup}\,,
	\end{align}
where
	\begin{align}
		I_n &= \iint _0^\infty dydy' yy' \mathrm{Ai}(\zeta _n + y)\mathrm{Ai}(\zeta _n + y') \nonumber \\
		&\times [\cos (\sqrt{-\zeta _n}(y-y'))+\mathrm{J}_0(\sqrt{-\zeta _n}(y-y'))]\,.
	\end{align}
\begin{figure}
	\centering
	\includegraphics[scale=0.53]{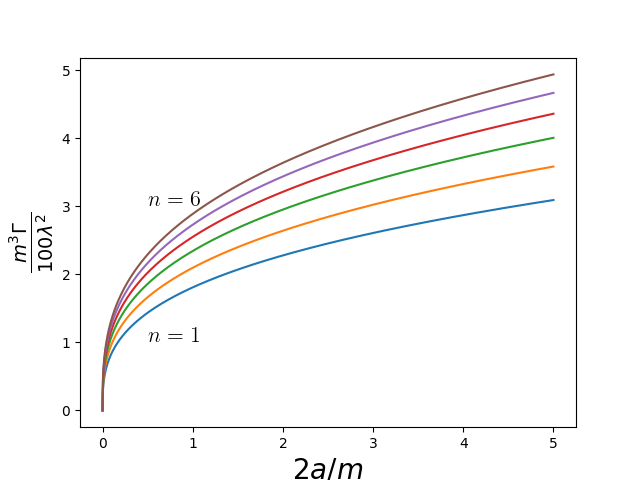}
	\caption{The decay path for various values of $n$, which is the index of the Airy function zero. It clearly differs from the flat space-time case, most notably by being discrete.}
	\label{fig:spec}
\end{figure}
 with $J_0$ a Bessel function of the first kind. The integral $I_n$ is only a numerical quantity and the essential parameter dependencies of the rate are included in its prefactor.  
 Note also, that as the metric space is static the decay rate is time independent.
 Figure 2 illustrates the decay rates for various parameter combinations.
	
	\subsubsection{\label{ssec:lszresult}LSZ reduction formula}
 
	For the case of a static spacetime, the formula \eqref{eq:lszred} can be vastly simplified. First of all, the combinations $\Lambda$ and $V$ of Bogoljubov coefficients are trivial. The outgoing modes will be plane waves, and incoming modes will be the Airy functions. We end up with the reduction formula
	\begin{align}
			\langle k_1k_2|\omega _p\rangle = i\mathcal{N}\int d^2y_1dy^2_2dx^2_1 \sqrt{-g_{y_1}g_{y_2}g_{x_1}} e^{ik_1\cdot y_1 + ik_2\cdot y_2}\nonumber \\
			\times \mathrm{Ai}(f(x,\omega)) e^{-i\omega _pt}\mathcal{K}_{y_1}\mathcal{K}_{y_2}\mathcal{K}_{x_1}\langle 0 | T(\psi (y_1)\psi (y_2)\phi (x_1))|0\rangle \label{eq:sm}
	\end{align}
	We now develop the Green's function as a series and use methods given in \cite{Birrell1979} to reduce the resulting expression. To first order
	\begin{align}
		&\langle 0 | T(\psi (y_1)\psi  (y_2)\phi  (x_1))|0\rangle \approx i\lambda\int dz \sqrt{-g_z} \nonumber  \\
		&\times \langle 0 |T \psi _0 (z)\psi _0(z)\phi _0(z)\psi _0(y_1)\psi _0(y_2)\phi _0 (x_1)|0\rangle\,,
	\end{align}
 where the subscript 0 denotes free field here. Then, generically according to Birrell and Taylor  (\cite{Birrell1979} Eq. 3.15),
	\begin{align}
		\mathcal{K}_x \tau (x,x_1,..x_n) =& -\frac{i}{\sqrt{-g_x}}\sum _{j=1}^n \delta (x-x_j) \nonumber \\ &\times \tau(x_1,\dots , x_{j-1}, x_{j+1},\dots ,x_n) \label{eq:red2a}\,,\\
		\tau (x,x_1,...x_n) =& \frac{\langle \text{out}, 0 | T\phi (x)...\phi (x_n)|\text{in},0\rangle }{\langle \text{out},0|\text{in}, 0 \rangle}\,.
	\end{align}
 
 First we calculate the first $\mathcal{K}_{x_1}$ operation in \eqref{eq:sm}.  According to equation \eqref{eq:red2a}, only one term survives,
	\begin{align}
		&\mathcal{K}_{x_1}\tau (x_1,y_1,y_2,z^\psi ,z^\psi,z^\phi ) \nonumber \\
		&= \frac{-i}{\sqrt{-g_{x_1}}}\bigg[ \delta (x_1-z)\tau (y_1,y_2,z^\psi ,z^\psi ) \bigg]\,, \label{eq:red1}
	\end{align} 
	while the rest of the terms are zero. Note, that the coordinates $z^\phi , z^\psi $ are all equal to the interaction point coordinate $z$ while we have labeled them just for for book keeping purposes.
	
	Next apply the $\mathcal{K}_{y_2}$ operation. Again, according to \eqref{eq:red2a}, 
	\begin{align}
		&\mathcal{K}_{y_2}\tau (y_1,y_2,z^\psi, z^\psi)\nonumber \\
		&=\frac{-i}{\sqrt{-g_{y_2}}} \bigg[2 \delta (y_2 - z)\tau (y_1,z^{\psi}) + \delta (y_2-y_1)\tau (z^\psi ,z^\psi )\bigg] \label{eq:red2},
	\end{align}
 where the factor two emerges from two separate but similar terms with same coordinate $z^\psi$.
	Finally we apply $\mathcal{K}_{y_1}$: obviously the only term left is
	\begin{align}
		&\mathcal{K}_{y_1}\tau (y_1,z^{\psi}) = \frac{-i}{\sqrt{-g_{y_1}}} \delta (y_1-z)\label{eq:red3}
	\end{align}
 as the operation to the latter term, $\mathcal{K}_{y_1}\delta (y_2-y_1)\tau (z^\psi,z^{\psi})$,
	gives effectively zero due to the Dirac delta function integration property $\int dx \partial _x \delta (x-y)f(x) = -\int dx\delta (x-y)\partial_x f(x)$.
	
	We may now insert eqs. \eqref{eq:red1}-\eqref{eq:red3} in to \eqref{eq:sm}, yielding
	\begin{align}
		\langle k_1k_2|\omega _p\rangle &=i\lambda \int d^2zd^2y_1d^2y_2d^2x_1 \sqrt{-g_{y_1}g_{y_2}g_{x_1}} e^{ik_1\cdot y_1 + ik_2\cdot y_2} \nonumber \\
		&\times \mathcal{N}\mathrm{Ai}(f(x,\omega)) e^{-i\omega _pt}\frac{-i}{\sqrt{-g_{x_1}}} \delta (x_1-z)\frac{-i}{\sqrt{-g_{y_2}}}\nonumber \\
		&\times2 \delta (y_2 - z)\frac{-i}{\sqrt{-g_{y_1}}} \delta (y_1-z).
	\end{align}
	After collecting terms, simplifications, and performing integrals over the Dirac delta functions we obtain
	\begin{align}
		\langle k_1k_2|\omega _p\rangle =&-2i\lambda\mathcal{N} \int d^2z \sqrt{-g_z}\,e^{ik_1\cdot z + ik_2\cdot z}\mathrm{Ai}(f(z,\omega)) \nonumber \\
		&\times e^{-i\omega _pt}
	\end{align}
	But this is exactly the same expression that we end up with in the added-up calculation --- the steps taken in the appendix apply to this integral just as well.
	
	In this case, it was easy to see the LSZ and added-up calculations give the same results. Let us now look at a time-dependent case.
	
 \subsection{\label{sec:quartet}Quartet creation in a time-dependent metric}
	
\begin{figure}

	\begin{tikzpicture}
		\begin{feynman}
			\node[crossed dot] (b);
			\vertex [] at (1.0,1.0)(f1) {$k_1$};
			\vertex [] at (1.5,0.5) (f3) {$k_2$};
			\vertex [] at (1.5, -0.5)(f4) {$k_3$};
			\vertex [] at (1.0, -1.0) (f2) {$k_4$};
			\vertex [] at (0.9,-1.90) {$(a)$};
			\diagram* {
				(b) -- [scalar] {(f1), (f2), (f3), (f4)},
			};
			
		\end{feynman}
	\end{tikzpicture}\hspace{1.0cm}
	\begin{tikzpicture}
		\begin{feynman}
			\node [crossed dot] at (0.0,0.8) (b);
			\vertex at (0.0,-0.75) (c);
			\vertex [] at (1.0,1.0)(f1) {$k_1$};
			\vertex [] at (1,0.6) (f3) {$k_2$};
			\vertex [] at (1.5, -0.0)(f4) {$k_3$};
			\vertex [] at (1.5, -1.5) (f2) {$k_4$};
			\vertex [] at (1.3, -0.75) (loop);
			\vertex [] at (0.9,-1.90) {$(b)$};
			\diagram* {
				(b) -- [scalar] {(f1), (f3)},
				(c) -- [scalar] {(f2), (f4)},
				(c) -- [scalar,half right, in=100, out=10] (loop) -- [scalar, half right, in=170, out=80] (c),
			};
			
		\end{feynman}
	\end{tikzpicture}\hspace{1.0cm}
	\begin{tikzpicture}
		\begin{feynman}
			\vertex(b);
			\vertex [] at (1.0,1.0)(f1) {$k_1$};
			\vertex [] at (1.5,0.5) (f3) {$k_2$};
			\vertex [] at (1.5, -0.5)(f4) {$k_3$};
			\vertex [] at (1.0, -1.0) (f2) {$k_4$};
			\vertex [] at (0.9,-1.90) {$(c)$};
			\diagram* {
				(b) -- [scalar] {(f1), (f2), (f3), (f4)},
			};
			
		\end{feynman}
	\end{tikzpicture}
	\caption{First order diagrams contributing to the equation \eqref{eq:lszquartet}. The crossed dot indicates particle creation by the space-time metric.  In diagram (a), all the 4 particles detected in the out-state were created by the metric. In (b), two were created by the metric, two by a self-interaction. In (c), all of them were created by self-interaction.}
	\label{fig:particlecreation}
	\end{figure}
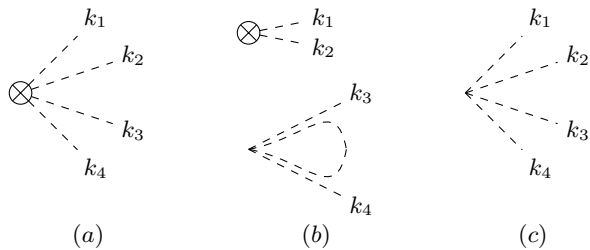

	In this case, we investigate the creation of massive scalar particles in a time-dependent metric, i.e. the matrix element $\langle k_1k_2k_3k_4|0\rangle$. Such a process exists e.g. in the Robertson-Walker metric with the $\phi ^4$ interaction \cite{Birrell1979b}. From the reduction formula we get:
	\begin{align}
		&\langle \text{out},k_1k_2k_3k_4|0, \text{in}\rangle \nonumber \\
		&= Q_{y_1}Q_{y_2}Q_{y_3}Q_{y_4}\langle 0,\text{out}|T \phi(y_1)\phi (y_2)\phi (y_3)\phi (y_4)|0,\text{in}\rangle + \nonumber\\
		&i\sum _{\sigma}V_{k_{\sigma (1)}k_{\sigma (2)}}Q_{y_{\sigma (3)}}Q_{y_{\sigma (4)}} \langle 0,\text{out}|T\phi(y_{\sigma (3)})\phi (y_{\sigma (4)})|0,\text{in}\rangle \nonumber \\
		&-(V_{k_1k_2}V_{k_3k_4}+V_{k_1k_3}V_{k_2k_4}+V_{k_2k_3}V_{k_1k_4})\langle 0,\text{out}|0,\text{in}\rangle \label{eq:lszquartet} 
	\end{align}
	We may interpret these as follows. The first term gives essentially the in-in amplitude, as discussed in \cite{Birrell1979b} for quartet creation: the self-interaction simply creates 4 particles and it corresponds figure \ref{fig:particlecreation}c. The following term, containing a sum over all possible combinations of $k_i$ indices, indicates a process by which two particles are generated by the space-time, two are created by self-interaction containing loop diagram to be renormalized (figure \ref{fig:particlecreation}b). Finally, the last term (figure \ref{fig:particlecreation}a) is a process in which all the four particles observed were generated by the space-time with no regard for the interaction, i.e. they are generated gravitationally. Hence, the amplitude is a sum of all the possible ways in which one could end up with four particles in the (asymptotically flat) out-vacuum. This seems to make sense: an observer with a detector may detect particles created for any of the above reasons.
	
	Thus, there is some interesting information that can be extracted from the in-out formalism but not from the in-in formalism: the likelihood of the different space-time particle creation processes. In the in-in formalism, particle creation by the space-time is taken in to account by performing the Bogoljubov transformation to the creation and annihilation operators. In the in-out formalism, to get the full density of particles created by the spacetime, one must take in to account possibilities $\langle k_1 k_2|$, $\langle k_1k_2k_3k_4|$, $\langle k_1k_2k_3k_4k_5k_6 |$, and so on (compare with the calculation in \cite{Birrell1979b}; in principle, an infinite number of particle pairs might be created at once. However, in-in formalism, one gets the density of particles by integrating over the square of the Bogoljubov coefficient $\beta _{ij}$.
	
	Let us have a look at the space-time generation of a particle pair for a very simple case: the 1+1 Robertson-Walker universe, for which calculations are performed in ref. \cite{Birrell1979b} using the in-in formalism (see also \cite{Birrell1982}). The line element is $ds^2 = dt^2 -a^2(t)dx^2$, which can be written in conformal form as $ds^2 = C(\eta )(d\eta ^2 -dx^2)$. Choosing the conformal factor as $C(\eta ) = A+B\tanh r \eta$, we make the following points:
	\begin{itemize}
		\item This universe is flat in the remote past and future.
		\item The amount of expansion is given by $2B$.
		\item The speed or "suddenness" of the expansion is proportional to $r$, which we call rapidity parameter.
	\end{itemize}

	For such a universe, the Bogoljubov coefficients are (details can be found in, for example, \cite{Birrell1982}):
	\begin{align}
		\alpha _{k} &= \bigg( \frac{\omega _{out}}{\omega _{in}} \bigg)^{1/2}\frac{\Gamma (1-(i\omega _{in})/r )\Gamma (-i\omega _{out}/r)}{\Gamma (-i\omega _+ /r) \Gamma (1+(i\omega _+ /r ))},\\
		\beta _{k} &= \bigg( \frac{\omega _{out}}{\omega _{in}} \bigg)^{1/2}\frac{\Gamma (1-(i\omega _{in})/r )\Gamma (i\omega _{out}/r )}{\Gamma (i\omega _- /r) \Gamma (1-(i\omega _+ /r ))},
	\end{align}
	with $\alpha _{kk'} = \alpha _k \delta _{kk'}$, $\beta _{kk'} = \beta _k \delta _{-kk'}$ and
	\begin{align}
		\omega _{in} &= (k^2+m^2(A-B))^{1/2},\\
		\omega _{out} &= (k^2+m^2(A+B))^{1/2},\\
		\omega _{\pm} &= \frac{1}{2}(\omega_{out}\pm \omega _{in}).
	\end{align}
	From these, one can infer
	\begin{align}
		|\alpha _k|^2 = \frac{\sinh ^2 (\pi \omega _+/r )}{\sinh (\pi \omega _{in}/r) \sinh (\pi \omega _{out}/r)},\\
		|\beta _k|^2 = \frac{\sinh ^2 (\pi \omega _-/r )}{\sinh (\pi \omega _{in}/r) \sinh (\pi \omega _{out}/r)}.
	\end{align}
	Using the in-in formalism, one can calculate the total number density of the particles created by space-time to the out-state, $\rho_{tot}=\langle 0, \text{in}|a_{out}^\dagger a_{out}|0,\text{in}\rangle = \int dk |\beta _k|^2$, by using the properties of the Bogoljubov transformation.
	
	Next we are now interested in computing the portion of this density caused by pair creation instead of quartet and higher order terms. We do this by applying the LSZ formula to the state $\langle k_1k_2|0\rangle$ and setting interactions to zero. Then
	\begin{align}
	\frac{	\langle k_1k_2, \text{out}|0, \text{in}\rangle}{\langle 0,\text{out} | 0,\text{in}\rangle } = iV_{k_1k_2} .\label{eq:amplspacecreat}
	\end{align}
	We can compute $V_{k_1k_2}$ by using the definition $V_{k_1k_2}=i\sum _k \beta ^*_{k_2k}\alpha ^{-1}_{kk_1} = i\int dk \beta ^*_{k_2k}\alpha ^{-1}_{kk_1}$. 
 
\begin{figure}
		\includegraphics[scale=0.55]{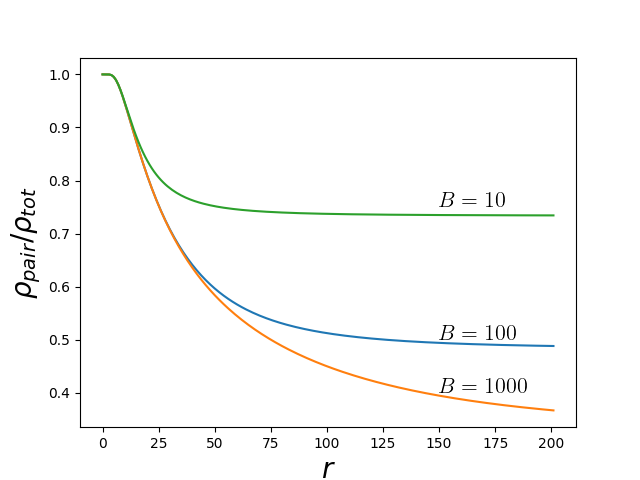}
		\caption{Fraction of space-time particle generation caused by a two-particle creation process, as a function of $\rho$, the rapidity of the expansion. The larger the post-expansion universe (determined by $B$), the more lower the fraction eventually becomes. All of the curves have a constant asymptotic value. \label{fig:twoparticle}}
	\end{figure}

 Investigating then $\langle k_1k_2, \text{out} |\hat{N}| k_1k_2, \text{out} \rangle$, which we do by using \eqref{eq:amplspacecreat} as the amplitude for a two-particle creation process, we find the density of particles created by a pair creation process
	\begin{align}
		\rho _{\text{pair}} = \int dk |\beta _k|^2|\alpha ^{-1}_k|^2,
	\end{align}
	which is compared numerically to the total number of particles created, the ratio $\rho_{pair}/\rho_{tot}$, in figure \ref{fig:twoparticle}. The numerics indicate that the proportion of four or more particle creation processes grow strongly along with growing rapidity parameter $\rho$ and amount of expansion $B$. Indeed, there are two asymptotics that are of special interest: the slow expansion limit near $r =0$, and the limit $r \sim \infty$ of sudden expansion. Near $r= 0$ the in-in formalism and the two-particle LSZ process give the same results as expected. The other asymptotic case $\rho \sim \infty$ is analytically too complicated for the LSZ case, but numerical calculations suggest that two particle creation process goes towards a non-zero asymptotic value. The asymptotic costant gets lower as $B$ gets larger.

	\section{\label{sec:discussion}Discussion}
	We have seen that the in-in and in-out formalisms are, as expected, equal in the case of static spacetimes, and even equal in some limited cases of time-dependent metrics when massless conformally coupled particles are involved. Though this can be proved in a general fashion, we have done it here by a direct calculation applying perturbation theory. We have also shown that certain conceptual issues can be clarified using the in-out, rather than the more customary in-in, formalism. The calculations here give a practical example to learn from for doing curved space LSZ calculations.
	
	The analysis shows that using the in-out formalism, we gain access to more information than with the in-in formalism, even though the physical observables calculated are the same. With the in-out formalism, we can see which channels are more or less likely sources of particle creation. In principle this allows us also to calculate the contribution of any creation channel at will, or cut off either low-energy or high-energy channels. Calculations will rapidly become more complicated as higher-order terms are taken in to account, as the in-out formalism contains more diagrams for each order of the series. Use of the in-out formalism could potentially lead to new insights in e.g. perturbative cosmological calculations; it might be interesting to see whether suppressing some channels would significantly change physical conclusions.

	\appendix
	\begin{widetext}
	\section{\label{app:addedup}Details of the added-up probability calculation}
	Starting with \eqref{reduced} and denoting the creation operators of the $\phi$-field with $a^\dagger$, and $b^\dagger$ for $\psi$, we note that $|\phi _p \rangle = a^\dagger _p|0\rangle$ and $\langle \psi _{k_1}\psi_{k_2}| = \langle 0 | b_{k_1}b_{k_2}$. Hence only a few terms survive in \eqref{reduced}. We obtain
\begin{align}
		|S^p_{k_1k_2}|^2  =&4\pi ^2 \lambda ^2  \delta (\omega _{k_1}+\omega _{k_2}-\omega _p) \int _0^\infty \int _0^\infty dx dx' (2a)^2xx'\mathcal{N}^2\text{Ai}\bigg( \frac{-\omega _p^2+2 a m^2 x}{2^{2/3} \left(a m^2\right)^{2/3}} \bigg)  \text{Ai}\bigg( \frac{-\omega _p^2+2 a m^2 x'}{2^{2/3} \left(a m^2\right)^{2/3}}\bigg) e^{i(k_1+k_2)(x'-x)}\nonumber\\
		=& (2a)^2 \mathcal{N}^2 4\pi ^2 \lambda ^2  \delta (\omega _{k_1}+\omega _{k_2}-\omega _p) \int _0^\infty \int _0^\infty dx dx' g(x,x')e^{i(k_1+k_2)(x'-x)},
	\end{align}
where we have denoted all the solely $x$-dependent terms as $g(x,x')$, which thus includes the Airy functions and vactor $x x'$.
	Since we are interested in the total decay rate, the momenta $k_1$ and $k_2$ must be integrated over. With the measure is $d\tilde{k} = \frac{dk}{\sqrt{2\omega _k}}$ the integrals are doable analytically. Firstly, the Dirac delta function makes the first integral trivial. The second integral splits in to four parts, according to whether k is negative or positive. We get the integral
\begin{align}
		\iint _{-\infty}^\infty d\tilde{k}_1d\tilde{k}_2 |S^p_{k_1k_2}|^2&=(2a)^2 \mathcal{N}^2 4\pi ^2 \lambda ^2\iint _{0}^\infty dxdx'g(x,x')\bigg[ \int _{-\omega _p}^{\omega _p} \frac{dk_2}{2\sqrt{|k_2|||k_2|-\omega _p|}}e^{i(|k_2|+k_2-\omega _p)(x'-x)} \nonumber \\
  &+\int _{-\omega _p}^{\omega _p} \frac{dk_2}{2\sqrt{|k_2||\omega _p-|k_2||}}e^{i(|k_2|-k_2+\omega _p)(x'-x)}\bigg] \nonumber \\
		&= (2a)^2 \mathcal{N}^2 4\pi ^2 \lambda ^2\iint _{0}^\infty dxdx'g(x,x')\pi \bigg( \cos (\omega _p (x'-x)) + J_0(\omega _p (x'-x)) \bigg).
	\end{align}
	Then we perform the variable change $y = \frac{2am^2x}{(2am^2)^{2/3}} = (2am^2)^{1/3}x, x = \frac{y}{(2am^2)^{1/3}}$. Because $\frac{-\omega _p ^2}{(2am^2)^{2/3}}$ is supposed to be a constant, an Airy function zero  $\zeta _n, \ n=1,\, 2,\, 3,\ \dots$, we obtain 
	\begin{align}
		g(y,y') &= \frac{yy'}{(2am^2)^{2/3}}\mathrm{Ai}\bigg( \zeta _n + y  \bigg)\mathrm{Ai}\bigg( \zeta _n + y'  \bigg),\\
		\cos (\omega _p (x'-x)) &= \cos \bigg(\omega _ p \bigg( \frac{y}{(2am^2)^{2/3}}-\frac{y'}{(2am^2)^{2/3}} \bigg)\bigg) = \cos (\sqrt{-\zeta _n}(y-y')),\\
		J_0(w_p (x'-x)) &= J_0(\sqrt{-\zeta _n}(y-y')).
	\end{align}	
We now get the desired result \eqref{eq:finaladded} directly by combining the three terms, and  dividing by the flux factor $4\omega _p^2$.
\end{widetext}

	\nocite{*}
		\bibliography{library}
		
	\end{document}